\documentclass[12pt]{article}
\pdfoutput=1
\usepackage[DIV13]{typearea}
\usepackage{graphicx}
\usepackage{amsmath}
\usepackage{amsfonts}
\usepackage{mathrsfs}
\usepackage{amssymb}
\usepackage{subfigure}
\usepackage{multicol}
\usepackage[usenames,dvipsnames]{color}
\usepackage{feynmp}
\usepackage{multirow}
\usepackage{array}
\usepackage{slashed}
\usepackage{units}
\usepackage{bbm}
\usepackage{framed}
\usepackage[left=.9in, right=.9in]{geometry}
\usepackage{makecell}
\usepackage{rotating}
\usepackage{cancel}
\usepackage{pifont}
\usepackage{tikz,feynmp}
\usepackage[normalem]{ulem}
\usepackage[font=small,labelfont=bf,margin=1cm]{caption}
\usepackage{cite}
\usepackage{hyperref}
\newcolumntype{L}[1]{>{\raggedright\let\newline\\\arraybackslash\hspace{0pt}}m{#1}}
\newcolumntype{C}[1]{>{\centering\let\newline\\\arraybackslash\hspace{0pt}}m{#1}}
\newcolumntype{R}[1]{>{\raggedleft\let\newline\\\arraybackslash\hspace{0pt}}m{#1}}

\newcommand{\blue}[1]{\textcolor{blue}{#1}}
\newcommand{\vev}[1]{\langle #1 \rangle}

\def\Ds{D\!\!\!\!/\,\,}
\newcommand{\paren}[1]{\left( #1 \right)}
\def\Ds{\slashed{D}}
\newcommand{\nn}{\nonumber}
\def\Lam{\Lambda}

\newcommand{\F}{\mathcal{F}}
\newcommand{\p}{\mathcal{P}}
\newcommand{\lam}{\lambda}

\newcommand{\dmu}{\partial_\mu}
\newcommand{\coma}{\, ,}
\newcommand{\VV}{\langle V_\mu V^\mu \rangle}

\newcommand{\ssh}{s^2_\varphi}

\newcommand{\sh}{s_\varphi}
\newcommand{\ch}{c_\varphi}

\newcommand{\order}[1]{\mathcal{O}\left(\frac{1}{\lambda^{#1}}\right)}
\def\De{\Delta}
\def\Ds{\slashed{D}}

\def\qr{\mathbf{q}_R}
\def\qri{\mathbf{q}_{iR}}
\def\qrj{\mathbf{q}_{jR}}
\def\barqrj{\bar{\mathbf{q}}_{jR}}

\title{
\flushright
{\small DFPD-2016/TH/17\\FERMILAB-PUB-16-471-T\\FTUAM-16-39\\IFT-UAM/CSIC-16-107\\}
\center
The linear-non-linear frontier for the Goldstone Higgs\\ 
}
\author{M.~B.~Gavela~$^{a)}$, K.~Kanshin~$^{b)}$, P.~A.~N.~Machado~$^{a,c)}$, and S.~Saa~$^{a)}$~\footnote{\href{mailto:gavela@uam.es}{\blue{\texttt{belen.gavela@uam.es}}}, \href{mailto:kanshin@pd.infn.it}{\blue{\texttt{kanshin@pd.infn.it}}}, \href{mailto:pmachado@fnal.gov}{\blue{\texttt{pmachado@fnal.gov}}}, \href{mailto:sara.saa@uam.es}{\blue{\texttt{sara.saa@uam.es}}}}
\\
\footnotesize $^{a)}$Departamento de F\'isica Te\'orica and Instituto de F\'{\i}sica Te\'orica, IFT-UAM/CSIC,\\
\footnotesize Universidad Aut\'onoma de Madrid, Cantoblanco, 28049, Madrid, Spain\\
\footnotesize $^{b)}$Dipartimento di Fisica e Astronomia `G.~Galilei', Universit\`a di Padova
\\
\footnotesize INFN, Sezione di Padova, Via Marzolo~8, I-35131 Padua, Italy
\\
\footnotesize $^{c)}$Theoretical Physics Department, Fermi National Accelerator Laboratory,\\
\footnotesize P.O. Box 500, Batavia, IL 60510, USA
}

\begin{document}

\maketitle 

\abstract{

 The minimal $SO(5)/SO(4)$ $\sigma$-model is used as a template 
for the ultraviolet completion of scenarios  in
 which the Higgs particle is a low-energy remnant of some high-energy dynamics, enjoying a (pseudo) Nambu-Goldstone boson ancestry. Varying the  $\sigma$ mass allows  
to sweep from the perturbative regime to the customary non-linear implementations.  The low-energy benchmark 
effective  non-linear Lagrangian for bosons and fermions is obtained, determining as well the operator coefficients  including  linear corrections.  At first order in the latter, three effective bosonic operators emerge which are independent of the explicit soft breaking assumed. The  Higgs couplings to vector bosons and fermions  turn out to be quite universal: the linear corrections are  proportional to the explicit symmetry breaking parameters.  Furthermore, we define an effective Yukawa operator which allows a simple parametrization and comparison of different heavy fermion ultraviolet completions. In addition, one particular fermionic completion is explored in detail, obtaining the corresponding leading low-energy fermionic operators.

\vspace{3cm}

\newpage

\tableofcontents

\section{Introduction}

The Higgs particle seems to be unnaturally light {\it if} there is new
particle physics at higher scales to which the Higgs may
couple. Barring a Copernican perspective on nature and the conclusion
that our generation has completed the discovery of particle physics of
the visible world, this puzzle --known as the ``electroweak hierarchy
problem"-- constitutes a pressing question.

The persistent absence of evidence for new resonances in the vicinity
of the electroweak scale calls for an in-depth exploration of beyond the Standard Model (BSM) theories 
which may separate and isolate the Higgs mass from the putative scale of
exotic BSM resonances. Pseudo-Nambu-Goldstone bosons
(PNGB) are naturally lighter than and decoupled from the rest of the
spectrum of their mother theory. This suggested decades ago that the
Higgs particle could be identified with the PNGB of some BSM high-energy theory~\cite{Kaplan:1983fs,
  Georgi:1984af, Dugan:1984hq}.  
  
In the initial proposal~\cite{Kaplan:1983fs} a global $SU(5)$ symmetry
was considered for the high-energy strong dynamics. Recent attempts
tend to start instead from a global $SO(5)$
symmetry~\cite{Agashe:2004rs, Contino:2006qr} at a high scale $\Lambda$, spontaneously broken to
$SO(4)$ and 
producing at this stage an ancestor of the Higgs particle in the form
of one of the resulting massless Goldstone bosons, with characteristic
scale $f$ and $\Lambda\le 4 \pi f$~\cite{Manohar:1983md}. The coset
$SO(5)/SO(4)$ represents the minimal possibility to interpret the
Higgs as a pseudo-Goldstone boson in the presence of a custodial
symmetry.  The explicit breaking of the global symmetry needed to
generate the electroweak scale $v\ne f$ and a mass for the Higgs
usually stems from soft couplings of the high-energy dynamics to the
Standard Model (SM) gauge bosons and fermions.
 
 Most of the literature on composite Higgs models based on (or
 containing) $SO(5)$ assumes from the start a strong dynamics and uses
 an effective non-linear formulation of the
 models~\cite{Contino:2006qr, Panico:2012uw, Carena:2014ria,
   Contino:2011np, Marzocca:2012zn, Redi:2012ha, Carmona:2014iwa,
   vonGersdorff:2015fta}, often denominated ``composite Higgs"
 scenario.  The ratio
\begin{equation}
  \label{xi}
  \xi\equiv \frac{v^2}{f^2}\,
\end{equation}
encodes the degree of non-linearity of a given model and is a measure
of the fine-tuning required to accommodate data.

A complete renormalizable model was instead  constructed in
Refs.~\cite{Barbieri:2007bh, Feruglio:2016zvt} (see also
Ref.~\cite{Contino:2011np, Alanne:2014kea} for related results), which in its scalar part is a
linear sigma model including a new singlet scalar $\sigma$. Furthermore, in Ref.~\cite{Feruglio:2016zvt} the procedure and first steps to obtain the non-linear effective Lagrangian were developed. Later work incides on interesting phenomenological consequences~\cite{Fichet:2016xvs,Fichet:2016xpw} and other aspects~\cite{Alanne:2016mmn,
  Buchalla:2016bse}.   That
minimal sigma model allows to gain intuition on the dependence on the
ultraviolet (UV) completion scale: it can be considered either as an
ultimate model made out of elementary fields, or as a renormalizable
version of a deeper dynamics, much as the linear $\sigma$
model~\cite{GellMann:1960np} is to QCD. Upon spontaneous breaking of
the electroweak symmetry, the Higgs and the $\sigma$ field mix, and
the resulting scalar sector is that of a system with two ``Higgs-like"
particles.\footnote{The heavier scalar is called  ``global Higgs" in
  Ref.~\cite{Fichet:2016xvs,Fichet:2016xpw}.}  The implications on
low-energy precision data and the expected signals at LHC have been
also developed in Refs.~\cite{Feruglio:2016zvt, Fichet:2016xvs, Fichet:2016xpw}, leading to a lower bound $m_\sigma \gtrsim 550$ GeV and interesting $\sigma$ decay channels into gauge bosons and $t\bar t$ pairs at LHC.
 
The linear-non-linear divide will be further explored here by varying the mass of
the extra scalar\footnote{When using further below polar coordinates
  the extra scalar will be dubbed $\rho$ as customary, see Sect.~2.}
$\sigma$: a light $\sigma$ particle corresponds to a weakly coupled
regime, while in the high mass limit the theory should fall back onto
a usual effective non-linear construction.  The effective low-energy
Lagrangian for non-linear realizations of electroweak symmetry
breaking has been determined previously, but the number of couplings
is very large in the most general case~\cite{Alonso:2012px,Buchalla:2013rka,Brivio:2013pma,Brivio:2016fzo}.  For the generic $SO(5)/SO(4)$ construction, a very reduced subset of those couplings, constituting a complete basis of bosonic operators,  was first established in Ref.~\cite{Alonso:2014wta}. This served  to generically parametrize 
scenarios of electroweak symmetry breaking in which the Higgs particle is a low-energy remnant of some
dynamics based on or containing $SO(5)$ as a global symmetry. 
 Here we will focus on the particular case of the minimal $SO(5)$ sigma model, leading to an even more reduced subset of operators  --the benchmark low-energy effective Lagrangian-- which are expected to be common to the non-linear limit of any construction containing the $SO(5)/SO(4)$ spontaneous breaking. We will consider here both bosonic and fermionic operators, though. The leading linear corrections and the leading dependence on the explicit $SO(5)$-breaking mechanism will be determined as well in this work. 

While the bosonic couplings should be universal, the fermionic part may instead be quite model-dependent. Many different
choices of exotic fermions have been explored in the literature,
mainly within non-linear realizations of the global symmetry (see
e.g. Ref.~\cite{Carena:2014ria}). Nevertheless, a common
characteristic of the so-called  ``partial compositeness" framework is that SM tree-level Yukawa couplings are forbidden by
the global symmetry, while instead  vertices coupling one or two heavy
exotic fermions to the Higgs field are allowed. Effective Yukawa
couplings between the SM and the Higgs field are thus induced at
low-energies, with a generic Seesaw-like pattern for the mass
generation of SM fermions, whose masses are then inversely
proportional to those of the heavy fermions.  We will analyze the
problem in two approaches:
\begin{itemize} 
 \item A rather-model independent one in which the field content of
   the SM is augmented exclusively by a singlet scalar within the
   minimal $SO(5)$ setup mentioned, while the leading
   phenomenological impact of heavy fermions is encoded in an
   effective Yukawa coupling of the SM fields that we will define. This effective operator will serve to
   parametrize and disentangle among different choices of BSM fermion embeddings.
 \item In a second step, a concrete choice for the heavy fermion
   representations will be considered~\cite{Feruglio:2016zvt}.
   This sector will be integrated out explicitly.
\end{itemize}
Note that in Ref.~\cite{Feruglio:2016zvt} we had already integrated out the specific BSM heavy fermion mentioned in this second step, although  leaving fully dynamical the scalar sector. That is,  the effective Lagrangian made out of SM fields plus the $\sigma$ particle was established.  It was also proposed there to next integrate out the latter, which is a straightforward procedure starting from that result. This task will be completed here in order to compare  with the case in which the order of integration of the heavy fields  (bosons versus fermions) is inverted. The resulting benchmark couplings will be also compared with those stemming from the procedure indicated in the first bullet above.

By furthermore keeping track of the linear and heavy-fermion corrections, the analysis will
provide candles to identify whether a renormalizable ultraviolet
completion exists in nature or alternatively an underlying
``composite" mechanism is at work at high-energy, analogous to QCD for the
chiral dynamics involving pions.

Note that the results may be relevant as well for other scenarios
based on global groups larger than $SO(5)$. Furthermore, a
Goldstone-boson parenthood for the Higgs is not exclusive of strong
interacting dynamical setups, but is also embedded in other
constructions such as ``little Higgs'' models, extra-dimensional
scenarios and others; our results will then apply as well to those
constructions.

The structure of the paper can be easily inferred from the Table of Contents.

\section{Model independent analysis}
Consider a Lagrangian 
\begin{equation}\label{eq:Complete_Lagrangian}
 \mathcal{L}=\mathcal{L}_s+\mathcal{L}_{\rm{f}}\, +\mathcal{L}_g\,,
\end{equation}
comprising, in its scalar sector $\mathcal{L}_s$, a linear sigma model which
  exhibits a global $SO(5)$ symmetry broken to $SO(4)$ and includes a 
  new scalar, $\sigma$, singlet under the SM gauge
  group~\cite{Feruglio:2016zvt}}
\begin{equation}\label{eq:Scalar_Lagrangian}
\mathcal{L}_s=\frac{1}{2}D_\mu \phi^T D^\mu \phi 
            - \lambda (\phi^T\phi-f^2)^2-\alpha f^3 \sigma + 2\beta f^2 H^\dagger H\,,
\end{equation}
where $\phi=(\widetilde{H},H,\sqrt{2}\sigma)/\sqrt{2}$ is a
$\mathbf{5}$-plet of $SO(5)$ encompassing the Higgs doublet degrees of
freedom $H$ in addition to $\sigma$. The terms which break
softly~\footnote{Additional soft breaking terms are possible, but only
  those proportional to $\alpha$ and $\beta$ are required to absorb
  one-loop counterterms and in this sense their inclusion leads to the
  minimal $\sigma$ model, see Ref.~\cite{Feruglio:2016zvt}.}  the
$SO(5)$ symmetry --proportional to $\alpha$ and $\beta$-- endow the
Higgs particle with a PNGB character, remaining naturally light as
long as $\alpha, \beta \ll \lambda$.  The embedding of the gauge group
$SU(2)_L\times U(1)_Y$ inside $SO(5)$ is purely
conventional. $\mathcal{L}_s$ contains as well the scalar interactions
with gauge bosons, with the $SU(2)_L\times U(1)_Y$ covariant
derivative given by
\begin{eqnarray}
D_\mu\phi=\paren{\dmu+ i g \Sigma_L^i W_\mu^i+ig'\Sigma_R^3B_\mu}\phi \,,
\label{covariant}
\end{eqnarray}
where $\Sigma_L^i$ and $\Sigma_R^i$ denote respectively the generators
of the $SU(2)_L$ and $SU(2)_R$ subgroups of the custodial $SO(4)$
group contained in $SO(5)$.  Both $h$ and $\sigma$ acquire a vacuum
expectation value (vev), leaving unbroken an $SO(4)'$ subgroup which
is rotated with respect to the group $SO(4)\approx SU(2)_L\times
SU(2)_R$ containing $SU(2)_L\times U(1)_Y$. $\mathcal{L}_g$ in
Eq.~(\ref{eq:Complete_Lagrangian}) encodes the kinetic terms for gauge
bosons.

Consider  now the fermion sector. A generic feature is that the
phenomenological constraints on partial compositeness require
additional vector-like fermions, which couple and act as mediators
among the SM fields. The exact form of the effective coupling is
model-dependent and varies according to how the SM fermions are
embedded in $SO(5)$.  We will obviate until
Sect.~\ref{explicit-heavy-fermions} the details of the heavy fermion spectrum, and use instead in this section a simplified --effective-- approach to the  dominant fermion-induced effects. 

The fermionic part of the Lagrangian
in Eq.~(\ref{eq:Complete_Lagrangian}) will be written as the sum of two
terms,
\begin{equation}\label{eq:Fermion_Lagrangian}
 \mathcal{L}_{\rm{f}}= \mathcal{L}_{{\rm{f}}, \rm SM}^{\rm kin} + \mathcal{L}_{\rm{f}}^{\rm Yuk}\,,
\end{equation}
where $\mathcal{L}_{{\rm{f}}, \rm SM}^{\rm kin}$ comprises the kinetic
terms for only SM fermions.  All what is needed here in addition is the
fact that, in frameworks akin to ``partial compositeness", the global
symmetry is explicitly broken by couplings between the SM fermions and
heavy exotic fermions, which are the source of: i) non-zero values for
the soft $SO(5)$ breaking parameters $\alpha$ and $\beta$ at one
loop, inducing a potential and mass for the Higgs particle; ii)
effective Yukawa couplings for the SM fermions and thus {\it the
  generation of SM fermion masses}.
 
 \begin{figure}[t]
\begin{center}
\begin{fmffile}{diag-eff}
\begin{equation*}
\begin{tikzpicture}[baseline=(current bounding box.center)]
\node { \fmfframe(0,0)(20,10){ 
      \setlength{\unitlength}{0.5mm}
	        \begin{fmfgraph*}(100,70)
		\fmfleft{i1}
		\fmfright{o1}
                		  \fmf{plain}{i1,v1}
                		  \fmf{plain}{v1,o1}
		\fmfblob{.25w}{v1}
		\fmffreeze
		  \fmfsurroundn{a}{10}
                  \fmf{dashes}{v1,b1,a2}
                  \fmf{dashes}{v1,b2,a4}
                  \fmf{dashes}{v1,a5}
                  \fmf{dashes}{v1,a7}
                  \fmf{dashes}{v1,b3,a8}
                  \fmf{dashes}{v1,b4,a10}
\fmflabel{$t_R$}{o1}
\fmflabel{$H$}{a2}
\fmflabel{$H$}{a4}
\fmflabel{$H$}{a5}
\fmflabel{$t_L$}{i1}
\fmflabel{$\sigma$}{a7}
\fmflabel{$\sigma$}{a8}
\fmflabel{$\sigma$}{a10}
\fmf{phantom,right,tension=0,tag=1}{b1,b2}
\fmf{phantom,right,tension=0,tag=2}{b3,b4}
\fmfposition
\fmfipath{p[]}
\fmfiset{p1}{vpath1(__b1,__b2)}
\fmfiset{p2}{vpath2(__b3,__b4)}
\fmfi{dots}{subpath (length(p1)/3,2*length(p1)/3) of p1}
\fmfi{dots}{subpath (length(p2)/3,2*length(p1)/3) of p2}
	        \end{fmfgraph*} }
    };
    \end{tikzpicture}
\end{equation*}
\end{fmffile}
\end{center}
\caption{\label{fig:fermion-mass-scheme} Schematic fermion mass
  operator at low scales with arbitrary insertions of the scalar
  fields.}
\end{figure}
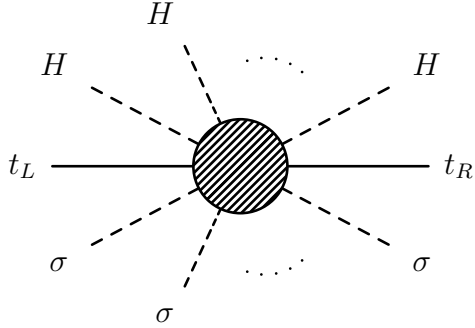

  \begin{table}
  \centering \renewcommand{\arraystretch}{2.1}
\begin{tabular}{|c| c|}
\hline
Fermion representation ($q_L$-$q_R$) & Yukawa interactions  $y_{\rm{f}}^0\mathcal{O}_{\rm Yuk}^{(n,m)}$ \\\hline
5-1, 5-10, 10-5 & $y\mathcal{O}_{\rm Yuk}^{(0,0)}$ \\\hline
5-5, 10-10, 14-10, 10-14, 14-1 & $y\mathcal{O}_{\rm Yuk}^{(1,0)}$ \\ \hline 
14-14 & $3y\mathcal{O}_{\rm Yuk}^{(1,0)}-2y'\mathcal{O}_{\rm Yuk}^{(1,1)}+8y'\mathcal{O}_{\rm Yuk}^{(3,0)}$ \\ \hline 
14-5 & $y\mathcal{O}_{\rm Yuk}^{(0,0)}+y'\mathcal{O}_{\rm Yuk}^{(2,0)}$  \\ \hline
5-14 & $y\mathcal{O}_{\rm Yuk}^{(0,0)}+y'\mathcal{O}_{\rm Yuk}^{(0,1)}-4y'\mathcal{O}_{\rm Yuk}^{(2,0)}$ \\ \hline
\end{tabular}
\caption{Yukawa operators corresponding to particular embeddings (see
  e.g. Ref~\cite{Carena:2014ria}) of a SM quark doublet $q_L$ and
  right-handed $q_R$ fermion (either up-type or down-type right-handed
  quark) into $SO(5)$. The coefficients $y$, $y'$ refer to distinct
  possible relative weights of $SO(5)$ invariant operators allowed by
  the models}
\label{table:yukawas}  
\end{table}
  
The schematic effective Yukawa coupling in the presence of the
$\sigma$ particle is presented in
figure~\ref{fig:fermion-mass-scheme}.  It follows that at low energies it is possible to write an
effective Yukawa Lagrangian  in terms of only the SM fermions, plus $h$ and $\sigma$, which respects
electroweak gauge invariance but not $SO(5)$ invariance,
\begin{equation}\label{eq:generic_yukawa_Lag}
  \mathcal{L}_{\rm{f}}^\text{Yuk} \equiv   - y_{\rm{f}}^0\, \mathcal{O}_{{\rm Yuk},{\rm{f}}}^{(n,m)}
  + \dots + {\rm h.c.}\,, 
  \end{equation}
where the constant $y_{\rm{f}}^0$ is a model-dependent
coefficient~\footnote{The superscript $0$ indicates that $y_{\rm f}^0$
  only encodes the leading contributions induced by the heavy
  fermionic sector.} and we define the effective Yukawa operator for a
given fermion $\rm{f}$ as
\begin{equation}\label{eq:generic_yukawa}
\mathcal{O}_{{\rm Yuk},\rm{f}}^{(n,m)} \equiv\bar{q}_L\widetilde{H}\, {\rm f}_R \left(\frac{\sigma}{f}\right)^n \left(\frac{2H^\dagger H}{f^2}\right)^m,
\end{equation}
with $\widetilde{H}\equiv-i\sigma^2 H^*$.  
The ellipses in Eq.~(\ref{eq:generic_yukawa_Lag}) refer to other SM
fermion operators and possibly extra model-dependent terms coming from
the heavy fermion sector.

In the literature of composite Higgs models the notation
MCHM$_{\rm{A-B-C}}$ is often used to indicate their fermion
composition, with $\rm{A, B, C}$ indicating the $SO(5)$ representation
in which the SM doublet $q_L$, up-type right-handed and down-type
right-handed fermions are embedded, respectively; else, when only one
subindex appears as in MCHM$_{\rm{A}}$ it is understood to be of the
type MCHM$_{\rm{A-A-A}}$. Table~\ref{table:yukawas} summarizes the
$\{n, m\}$ parameter values for different models.~\footnote{Models
  with spinorial $SO(5)$ embeddings,
  e.g. MCHM$_4$~\cite{Agashe:2004rs}, are phenomenologically
  excluded in particular in view of $Z\to b
  \bar{b}$ data.\cite{Contino:2006qr}}

Eq.~(\ref{eq:generic_yukawa}) assumes that a given fermion mass corresponds to a single set  of $\{n, m\}$ values. This is often the case; for instance, the top and bottom Yukawa couplings in the  MCHM$_{5-1-1}$
model~\cite{Feruglio:2016zvt} correspond to 
$\mathcal{O}_{{\rm  Yuk}}^{(0,0)}$, while in the MCHM$_5$ scenario they both correspond to $\mathcal{O}_{{\rm
    Yuk}}^{(1,0)}$ (see e.g. Ref.~\cite{Panico:2015jxa}). Notice that, for these cases with a single Yukawa operator, 
 the global coefficients and suppression scales in Eqs.~(\ref{eq:generic_yukawa_Lag})-(\ref{eq:generic_yukawa}) 
 are constrained by the fermion masses and therefore do not
constitute any additional model dependence.

Nevertheless, in some scenarios a given fermion mass results instead
from combining several operators of the type in
Eq.~(\ref{eq:generic_yukawa}) with different $\{n,m\}$ values. The
procedure derived can be easily extended to encompass it. A
model-dependence remains then in the relative size of the $y$ and $y'$
weights in Table~\ref{table:yukawas}. An example is the
MCHM$_{14-14-10}$ scenario~\cite{Carena:2014ria} in which different
sets of $\{n, m\}$ values are involved in generating the top mass,
while the bottom mass only requires set $\{n,m\}=\{1,0\}$.  The cases
of single and of multiple Yukawa operators contributing to a given
mass will be further considered explicitly below. We focus in
what follows on the top Yukawa coupling unless otherwise explicitly stated, while the conclusions to be
obtained are easily generalized to all light fermions.

\vspace{1cm}

\subsubsection*{Polar coordinates}
Armed with the tools described, it is quite straightforward to derive
 the benchmark bosonic Lagrangian as well as the leading couplings
involving fermions.  To this aim, it is convenient to rewrite the
scalar degrees of freedom in polar coordinates,
\begin{eqnarray}
 \sigma&\equiv&\rho\, c_\varphi\,, \\
 H&\equiv&\frac{1}{\sqrt{2}}\,\rho\,\,U\,s_\varphi\,,
\end{eqnarray}
with $c_\varphi \equiv \cos\varphi/f$, $s_\varphi\equiv
\sin\varphi/f$, and $U(x) \equiv \exp\{2 i \Pi(x)/f\}$, where $\Pi(x)$
denotes the Goldstone matrix corresponding to the longitudinal
components of the electroweak gauge bosons.  In this notation the
scalar Lagrangian in Eq.~(\ref{eq:Complete_Lagrangian}) reads
\begin{equation}\label{eq:lagrangian-scalar-polar}
  \mathcal{L}_s=\frac{1}{2}\partial_\mu\rho\,\partial^\mu\rho 
        + \frac{\rho^2}{2f^2}\left[ \partial_\mu\varphi\,\partial^\mu\varphi 
                                  - \frac{f^2}{2}s_\varphi^2 \vev{V_\mu V^\mu}\right]  
        -\lambda(\rho^2-f^2)^2 - \alpha f^3 \rho\, c_\varphi + \beta f^2\rho^2 s_\varphi^2, 
\end{equation}
where $\vev{\,}$ denotes the trace and $V_\mu \equiv (D_\mu
U)U^\dagger$ as is customary.  The effective top Yukawa operator in
Eqs.~(\ref{eq:generic_yukawa_Lag}) and (\ref{eq:generic_yukawa}) is
then given by
\begin{equation}\label{eq:fermion-yukawa}
 y_t^0\, \mathcal{O}_{{\rm Yuk},t}^{(n,m)}=\frac{y_t^0}{\sqrt{2}}(\bar{q}_LU P_+ \mathbf{q}_R)\rho\left(\frac{\rho}{f}\right)^{n+2m}c_\varphi^n s_\varphi^{2m+1}\,,
\end{equation}
where the right-handed SM fermions have been gathered in a formal
doublet $\qr\equiv(t_R,\,\,b_R)$, with $P_+\equiv{\rm
  diag}(1,\,\,0)$ ($P_-={\rm diag}(0,\,\,1)$) being a projector onto
the up-type (down-type) right-handed SM fermions.

The $\rho$ and $\varphi$ fields will develop vevs, 
\begin{equation}
  \rho\to\rho+\vev{\rho},\qquad \varphi \to h + \vev{\varphi}\,,
  \label{vevs-polar}
\end{equation}
where at the minimum of the potential the $\varphi$ field corresponds to 
\begin{equation}\label{cos-min}
 \cos\left(\frac{\vev{\varphi}}{f}\right)=-\frac{\alpha}{2\beta}\,{\left(1 +\frac{\beta}{2\lambda}\right)^{-1/2}}\,.
\end{equation}
The connection between the vevs of the fields in the linear and polar
parametrizations is
\begin{equation}
\vev{\rho} = \sqrt{\vev{\sigma}^2+2\vev{H}^2 },
	\qquad\vev{\varphi}=\,f\,\tan^{-1}\paren{\frac{\sqrt{2}\vev{H}}{\vev{\sigma}}}.
\end{equation}
The scalar resonance, which in the linear parametrization is
customarily denoted $\sigma$, is traded by $\rho$ in the polar
parametrization, with $m_\rho=m_\sigma$ exactly as expected for a physical observable, while the Higgs
resonance  $h$ corresponds now to the excitation of the $\varphi$
field, see Eq.~\ref{vevs-polar}.

Finally, as the pure gauge Lagrangian $\mathcal{L}_g$ and the weak coupling to fermions are not modified, the coefficient of the $W_\mu$ mass term in Eq.~(\ref{eq:lagrangian-scalar-polar}) allows to identify the electroweak scale $v$ in terms of the Lagrangian parameters:
\begin{equation}
\label{EWscale}
v^2=\vev{\rho}^2\, \sin^2\left(\frac{\vev{\varphi}}{f}\right)\,.
\end{equation}

\boldmath
\subsubsection*{Expansion in $1/\lambda$}
\unboldmath
The scalar
quartic coupling $\lam$ can be conventionally traded by the $\rho$
mass, given by $m_\rho^2\simeq 8 \lambda f^2$ for negligible $\alpha$ and
$\beta$, see Ref.~\cite{Feruglio:2016zvt} and further below; the non-linear model would be recovered in the limit
$m_\rho\gg f$, that is $\lambda\to\infty$. Varying the $\rho$ mass (that is, $\lambda$) allows to sweep from the regime of
 perturbative ultraviolet completion to the non-linear one assumed in 
 models  in
 which the Higgs particle is a low-energy remnant of some strong
 dynamics. We will explore this limit next.

The exact equation of motion for $\rho$ reads
\begin{align}
  \square \rho - &\frac{\rho}{f^2}\left[ \partial_\mu\varphi\,\partial^\mu\varphi 
    - \frac{f^2}{2}s_\varphi^2 \vev{V_\mu V^\mu}\right]  
  +4\lambda\rho(\rho^2-f^2)
  + \alpha f^3 c_\varphi - 2\beta \rho f^2 s_\varphi^2\nonumber\\
  \label{eq:eom-rho}
  &\quad\qquad+ (n+2m+1)\left(\frac{y_t^0}{\sqrt{2}}\bar{q}_LU P_+ \qr+{\rm h.c.}\right) c_\varphi^n s_\varphi^{2m+1}\left(\frac{\rho}{f}\right)^{n+2m}=0,
\end{align}
where $\square\equiv\partial_\mu\,\partial^\mu$.  In a $1/\lambda$ expansion, the $\rho$ field can be expressed as
$$\rho \equiv \rho_0 + \rho_1/\lambda + \rho_2/\lambda^2+\dots$$
where the leading terms are given by 
\begin{align}
  \rho_0 &= f,\\ 
  \label{eq:expressions-rho}
  \rho_1 &= \frac{f}{4}
  \Big[\frac{1}{2f^4}\partial_\mu \varphi\,\partial^\mu\varphi
    -\frac{1}{4f^2}\vev{V_\mu V^\mu}s_\varphi^2 - \frac{1}{2}\alpha c_\varphi + \beta s_\varphi^2\nonumber\\
    & \hspace{3.5cm}- \frac{(n+2m+1)}{2f^3}\left(\frac{y_t^0}{\sqrt{2}}\bar{q}_LU P_+ \qr+{\rm h.c.}\right) c_\varphi^n s_\varphi^{2m+1}\Big],
\end{align}
and subsequent ones can be written as polynomial functions of $\rho_1$.  Substituting
these in Eq.~(\ref{eq:lagrangian-scalar-polar}) yields the $1/\lambda^n$ Lagrangian
corrections,
\begin{equation}
\mathcal{L} = \mathcal{L}_0 + \mathcal{L}_1/\lambda + \mathcal{L}_2/\lambda^2+\dots\,,
\label{L_expanded}
\end{equation}
where the different terms in this equation are given by
\begin{align}\label{eq:L_0}
  \mathcal{L}_0&=\frac{1}{2}\partial_\mu \varphi\,\partial^\mu\varphi 
  -\frac{f^2 }{4}\vev{V_\mu V^\mu}s_\varphi^2 - \alpha f^4 c_\varphi + \beta f^4 s_\varphi^2 - f\,\frac{y_t^0}{\sqrt{2}}\bar{q}_LU P_+ \qr c_\varphi^n s_\varphi^{2m+1},
\\\label{eq:L_1}
\mathcal{L}_1&=4f^2 \rho_1^2, \\\label{eq:L_2}
\mathcal{L}_2&=\frac{1}{2}\Bigg\{(\partial_\mu\rho_1)^2 +\bigg[\alpha f^2c_\varphi\\
                &\hspace{3.2cm}
                +\left(1-(n\!+\!2m)^2\right)\frac{1}{f}\left(\frac{y_t^0}{\sqrt{2}}\bar{q}_LUP_+\qr+{\rm h.c.}\right)c_\varphi^ns_\varphi^{2m+1})\bigg]\rho_1^2\Bigg\}\,. \nonumber
\end{align}
 $\mathcal{L}_0$ coincides with the leading-order Lagrangian for the scalar sector of the minimal composite Higgs
model~\cite{Agashe:2004rs}, as expected.  The expressions obtained for 
$\mathcal{L}_1$ and $\mathcal{L}_2$ are remarkably compact and a similar pattern holds for higher orders in $1/\lambda$.

The maximum number of derivatives of $\mathcal{L}_n$ is $2+2n$,
although not all $2+2n$ derivative operators are generated at order
$n$. This is as foreseen, as for large $\lambda$ the non-linear regime
is approached and ordering the operators by their $1/\lambda$
dependence does not coincide with the ordering given by mass
dimensions.  The ordering in which the operators appear is akin to the
power counting of non-linear Higgs effective
theory~\cite{Manohar:1983md, Cohen:1997rt, Luty:1997fk, Gavela:2016bzc}.

Eqs.~(\ref{eq:expressions-rho}) and (\ref{eq:L_1}) suggest interesting
correlations between operators involving the Higgs boson, gauge bosons
and fermions. In particular, operators such as 
$(\partial_\mu h)^2\bar\psi\psi$ or $\vev{V_\mu
  V^\mu}\bar\psi\psi$, where $\psi$ denotes a generic fermion, are
weighted by the fermion mass and also bear a dependence on the SM
embedding into $SO(5)$, parametrized by  the set $\{n,m\}$  in
Eq.~(\ref{eq:fermion-yukawa}).  From those equations emerges the low-energy effective Lagrangian in terms of SM fields
 at a given order in $1/\lambda$,
 \begin{equation}
  \mathcal{L}_{\rm eff}=\mathcal{L}_g + \mathcal{L}_{{\rm{f}}, \rm SM}^{\rm kin}
  +\sum_i \p_i \,\F_i(\varphi),
\end{equation}
where the first two terms in the right-hand side contain respectively the kinetic
terms for gauge bosons and fermions as in Eqs.~(\ref{eq:Complete_Lagrangian}) and (\ref{eq:Fermion_Lagrangian}), and the index $i$ runs over all
operator labels and coefficient functions $\F_i(\varphi)$ in Table~\ref{table:operators}.
The table collects all couplings corresponding to
two and four ``derivatives", where plain derivatives and gauge
  boson insertions are counted with equal weight, as they come
  together in the covariant derivative. The notation/basis for the purely bosonic
operators was chosen according to Ref.~\cite{Alonso:2014wta} to
facilitate the comparison with a model-independent approach. From this
tree-level analysis we draw the following conclusions:
\begin{itemize}
\item Up to first order in the linear corrections, the benchmark
  effective Lagrangian is determined to be composed of ten operators,
  five of them bosonic and the rest fermionic~\footnote{For fermionic operators only the generic Lorentz and flavor structure are explicited, being trivial their decomposition in terms of  different  flavors.}
   including that
  responsible for Yukawa couplings. The coefficients of those operators are not free but intimately correlated by the coefficient functions explicitly determined in this work, and shown in the table. 
\item Among the couplings  which first appear at
  $\mathcal{O}(1/\lambda)$, three bosonic operators are singled out
  in the SO(5)-invariant limit ($\alpha=\beta=0$, massless SM
  fermions): $\p_6$, $\p_{20}$ and $\p_{DH}$.  The two latter ones
  involve multiple Higgs insertions and are out of present
  experimental reach while the strength of $\p_6$, which involves vertices with four gauge bosons, is already tested directly by data, although the present sensitivity is  very 
weak~\cite{Aad:2014zda,Aad:2016ett}.\footnote{ These bounds can be translated for instance 
in $m_\rho\gtrsim 70$ GeV for $f\approx 600$ GeV.}
\item {Operators involving SM fermions have an implicit dependence
  on the symmetry-breaking terms in the Lagrangian -- they are
  weighted by the fermion masses in a pattern alike to that of the Minimal Flavour
  Violation setup~\cite{Chivukula:1987py,D'Ambrosio:2002ex,Grinstein:2010ve}. Most interestingly, the corresponding $\F_i(\varphi)$ coefficients, written as a function  of the $\{n,m\}$ parameters, allow to differentiate the expected impact of different fermionic ultraviolet completions in the literature.
\item All operators derived from
  Eqs.~(\ref{eq:L_0})-(\ref{eq:L_2}) are at most four-derivative ones,
  and they are all  shown in table~\ref{table:operators}, including the only one appearing at 
  $\mathcal{O}(1/\lambda^3)$.}
\item The gauge field dependence is present only through powers of $\vev{V_\mu
  V^\mu}$, consistent with its exclusively scalar covariant derivative
  origin, see Eqs.~(\ref{eq:Scalar_Lagrangian}) and
  (\ref{eq:lagrangian-scalar-polar}).  Other Lorentz contractions such
  as $\vev{V_\mu V_\nu}^2$ would be loop-induced, and thus expected to be subleading.
\item  The scalar functions
  $\F_i(\varphi)$ obtained as operator weights of bosonic
  couplings are in agreement with those derived in
  Ref.~\cite{Alonso:2014wta} in the $SO(5)$ invariant limit,
for the subset of operators identified here as benchmarks, see their Eqs.~(2.5)-(2.8). Table~\ref{table:operators} provides in addition the leading
  deviations due to the presence of explicit $SO(5)$-breaking
  parameters $\alpha$ and $\beta$.
\end{itemize}

\begin{table}
  \centering \renewcommand{\arraystretch}{2.1}
\begingroup\small
\begin{tabular}{|c| c| c|c| }
\hline
&Operator &  $\F_k(\varphi)$  &  $1/\lambda^n$ \\
\hline 
\hline 
$\p_H$ & $\dfrac{1}{2}(\dmu h)^2$ &  
	$1-\dfrac{1}{4\lambda}(\alpha\ch-2\beta\ssh)$  &  0 \\  
\hline 
$\p_C$ & $-\dfrac{v^2}{4}\VV$&
	$ \dfrac{1}{\xi}\left [1-\dfrac{1}{4\lam}
				\paren{\alpha\ch-2\beta\ssh}\right ]s_\varphi^2$ &   0 \\
\hline
$\p_{\rm Yuk}$ &  $ v\,\bar q_{iL} U P_\pm \qri +{\rm h.c.}$ & 
$-\dfrac{y_i^0}{\sqrt{2\,\xi}}c_\varphi^ns_\varphi^{2m+1}\left(1-\dfrac{n\!+\!2m\!+\!1}{8\lambda}(\alpha c_\varphi -2\beta s_\varphi^2)\right)$&  0 \\
\hline\hline
$\p_{DH}$&  $\dfrac{1}{v^4}(\dmu h)^4$
	&$ \dfrac{\xi^2}{16 \lam}$  &  1 \\ 
\hline
$\p_6$ &  $\VV^2$& $\dfrac{s^4_\varphi}{64 \lam}$ &  1 \\
\hline
$\p_{20}$ &  $\dfrac{1}{v^2}\VV(\partial_\nu h)^2$& 
	$-\dfrac{\xi}{16\lam}s_\varphi^2$    &  1 \\
\hline
$\p_{qH}$ &  $\dfrac{1}{v^3}(\partial_\mu h)^2\bar q_{iL} U P_\pm \qri+{\rm h.c.}$ & 
$-\dfrac{y_i^0}{\sqrt{2}}\xi^{3/2}\left(\dfrac{n\!+\!2m\!+\!1}{8\lambda}\right)c_\varphi^ns_\varphi^{2m+1}$ &  1 \\
\hline
$\p_{qV}$ &  $\dfrac{1}{v}\vev{V_\mu V^\mu}\bar q_{iL} U P_\pm \qri+{\rm h.c.}$ & 
$\dfrac{y_i^0}{\sqrt{2}}\sqrt{\xi}\left(\dfrac{n\!+\!2m\!+\!1}{16\lambda}\right)c_\varphi^ns_\varphi^{2m+3}$ & 1  \\
\hline
$\p_{4q}$ &  $\dfrac{1}{v^2}(\bar q_{iL} U P_\pm \qri)(\bar q_{jL} U P_\pm \qrj)+{\rm h.c.}$ & $(2-\delta_{ij})y_i^0y_j^0\xi\dfrac{(n\!+\!2m\!+\!1)^2}{32\lambda}c_\varphi^{2n}s_\varphi^{4m+2}$&  1 \\
\hline
$\p_{4q'}$ &  $\dfrac{1}{v^2}(\bar q_{iL} U P_\pm \qri)(\barqrj P_\pm U^\dagger  q_{jL})+{\rm h.c.}$ & $(2-\delta_{ij})y_i^0y_j^{0}\xi\dfrac{(n\!+\!2m\!+\!1)^2}{32\lambda}c_\varphi^{2n}s_\varphi^{4m+2}$&  1 \\
\hline
\hline
$\p_7$ &  $\dfrac{1}{v}\VV(\Box h)$& $\sqrt{\xi} \left [
		\dfrac{1}{128 \lam^2}\paren{\alpha+4\beta\ch}s^3_\varphi
		\right ]$& 2  \\
\hline
$\p_{\Delta H}$ &  $\dfrac{1}{v^3}(\dmu h)^2(\Box h)$&  $-\xi^{3/2}\left[
		\dfrac{1}{64 f^3\lam^2}\paren{\alpha+4\beta\ch}\sh
		\right]$& 2  \\
\hline
\hline
$\p_{\Box H}$ &  $\dfrac{1}{v^2}\paren{\Box h}^2$& 
		$\mathcal{O}\paren{\dfrac{1}{\lam^3}}$ & 3  \\
\hline
\end{tabular}
\endgroup
\caption{Effective operators before electroweak symmetry breaking,
  including two and four derivative couplings, together with their
  coefficients up to their first corrections in the  $1/\lambda$ expansion.  The bosonic
  contributions from $SO(5)$ breaking contributions ($\alpha\ne 0$
  and/or $\beta\ne 0$) are also shown. The right-hand column indicates the order in  $1/\lambda$ at which a given couplings first appears.  The Higgs field $h$ is defined as the
  excitation of the field $\varphi$, see Eq.~(\ref{vevs-polar}). 
}
\label{table:operators}  
\end{table}

\subsection {Impact on Higgs observables}
\subsubsection*{Bosonic sector}
From Eqs.~(\ref{eq:L_0}) and (\ref{eq:L_1}), the potential at order
$1/\lambda$ reads
\begin{equation}
  \frac{V}{f^4} =\alpha c_\varphi - \beta s_\varphi^2 
     - \frac{1}{16\lambda}\left(\alpha c_\varphi -2\beta s_\varphi^2\right)^2 + \order{2}\,,
\end{equation}
with minimum at $
\cos\left(\frac{\vev{\varphi}}{f}\right)\simeq-\frac{\alpha}{2\beta}\left(1-\frac{\beta}{4\lambda}\right)$,
see Eq.~(\ref{cos-min}).  The kinetic
energy of the physical Higgs excitation $h$  (see Eq.~(\ref{vevs-polar})) gets then a correction given by
\begin{equation}
  \frac{1}{2}\left(1+\frac{\beta}{2\lambda}\right)\partial_\mu h\,\partial^\mu h\,,
\end{equation}
which is reabsorbed by a field redefinition
\begin{equation}\label{eq:Zh}
h\to (1+Z_h)\, h\,, \qquad {\rm{with}} \qquad  Z_h = -\frac{\beta}{4\lambda}\,.
\end{equation}
{\bf Renormalization.} The four independent parameters of the scalar
Lagrangian Eq.~(\ref{eq:Scalar_Lagrangian}), $f$, $\lambda$, $\alpha$
and $\beta$, can be expressed in terms of the following
observables~\cite{Feruglio:2016zvt}:
\begin{equation}
G_F\equiv(\sqrt{2} v^2)^{-1},\qquad m_h,\qquad \kappa_V\,, \qquad m_\rho\,
\label{renorm-scalar}
\end{equation} 
with the Fermi constant $G_F$ as measured from muon decay, while $\kappa_V$ can be extracted 
 from deviations of the Higgs couplings to two gauge bosons, for instance 
\begin {equation}
  \Gamma(h\to W W^*) \equiv \Gamma_{\rm SM}(h\to WW^*)\,\kappa_V^2\,.  \nn \\
\end{equation} 
$m_h$ is determined from the Higgs pole mass and $m_\rho$ could in
turn be determined from future measurements of the $\rho$ mass,
identifying them respectively with the light and heavy mass
eigenvalues of the scalar sector~\cite{Feruglio:2016zvt}
\begin{equation} 
m^2_{\rm heavy, light}=
	4\lambda f^2\left\{ \left(1+\frac{3}{4}\frac{\beta}{\lambda} \right) \pm 
	\left[1+ \frac{\beta}{2 \lambda}\left(1 +\frac{\alpha^2}{2 \beta^2} 
	+\frac{\beta}{8 \lambda}\right) \right]^{1/2} \right\}\,, 
\label{mhsigma}
\end{equation} 
where the plus sign refers to the heavier eigenstate.  Assuming the
$SO(5)$ explicit breaking to be small, ${|\beta|}/{4\lam}\ll1$, the
mass eigenvalues read
\begin{align}
m_\rho^2=&~ 8\lambda f^2 + 2\beta (3 f^2 - v^2) +\mathcal{O}\paren{\frac{\beta}{4\lambda}}\,,\label{mrho}\\
m_h^2=&~ 2\beta v^2+\mathcal{O}\paren{\frac{\beta}{4\lambda}}\,.
\label{mh}
\end{align}
with the measured value of $m_h$ implying $\beta\simeq 0.13$. In the
non-linear limit $\lambda\to\infty$ the $\rho$ field decouples from
the spectrum and the scalar sector would depend on just three
renormalized parameters.  It is now possible to foresee the impact of
the linear corrections in terms of mass dependence. Precisely because
the large $\lambda$ and $m_\rho$ limits are in correspondence,
dimensional arguments suggest the equivalence
\begin{equation}
  \frac{1}{\lambda} \qquad\Rightarrow \qquad \frac{m_h^2}{m_\rho^2}\,\simeq \frac{\beta \xi}{4\lambda}\,,
  \label{linear-correction}
\end{equation}
as expansion parameter, see Eqs.~(\ref{mrho}) and (\ref{mh}).  In
other words, the linear corrections are expected to be proportional to
the two small parameters $\beta$ and $\xi$ and thus doubly suppressed.

Extending the renormalization scheme to the gauge sector, we choose the two extra
observables needed to be the mass of
the $Z$ boson and the fine structure constant,
\begin{equation}
M_Z, \qquad
\alpha_{em}=\frac{e^2}{4\pi}\,,
\label{renorm-scalar}
\end{equation}
with $M_Z$ and $ \alpha_{em}$ as determined from Z-pole mass
measurements and from Thompson scattering, 
respectively~\cite{Agashe:2014kda}. 
 In terms of the ensemble of renormalized parameters discussed above, predictions can now be made. 
For instance  the relation between the gauge boson masses remains the same than that in the SM, 
\begin{equation}
M_W=\cos\theta_W M_Z \coma 
\end{equation}
where the weak angle is given at tree-level by
\begin{equation}
\sin^2\theta_W=\frac{1}{2}\paren{1-\sqrt{1-\frac{4\pi \alpha_{em}}{\sqrt{2}G_FM_Z^2}}}\,.
\end{equation}   
As another example, the prediction for the Higgs$\to ZZ$ width is modified with respect to the SM expectation by
\begin{equation}
\Gamma(h\to ZZ^*) =\Gamma_{\rm SM}(h\to ZZ^*)\,\kappa_V^2\,. 
\end{equation}

A generic expectation is the departure of $\kappa_V$ from $1$.
Indeed, the coupling between the Higgs and the gauge bosons which
stems from the Lagrangian Eqs.~(\ref{L_expanded})-(\ref{eq:L_2}) at
order $1/\lambda$ is that encoded in the operator $\p_C$ in
Table~\ref{table:operators} and reads
\begin{equation}
  \mathcal{L}_{hVV} = -\left(\frac{1}{2}\sqrt{1-\xi}
  +\frac{\beta}{8\lambda}\frac{(2-\xi)\xi}{\sqrt{1-\xi}} \right)
  \vev{V_\mu V^\mu} vh\,,
\end{equation}
or in other words
\begin{equation}
  \kappa_V = \sqrt{1-\xi} + \frac{\beta\xi}{2\lambda}\frac{(1-\xi/2)}{\sqrt{1-\xi}}\,.
    \label{eq:kappa_V}
\end{equation}
Assuming for illustrative purposes $\mathcal{O}(\xi)\sim
\mathcal{O}(1/\lambda)$ and expanding up to second order in these
parameters, the result simplifies to
\begin{equation}
  \kappa_V\simeq \sqrt{1-\xi}+\frac{\beta\xi}{2\lambda}.
  \label{eq:kappa_V_expanded}
\end{equation}
The first term on the right-hand side of this equation is the
well-known correction present in non-linear
scenarios~\cite{Agashe:2004rs}, while the second term encodes the
linear correction linked to the scale of ultraviolet completion, which
in terms of physical parameters we predict to be given by
\begin{equation}
  \kappa_V^2\simeq 1-\xi +4\frac{m_h^2}{m_\rho^2}\,,
\end{equation}
where Eq.~(\ref{linear-correction}) has been used. 
Higher order corrections are expected to be
very small, as they will stem from operators with at least 4
derivatives. For instance, the first extra tree-level contribution to
$\kappa_V$ is the $1/\lambda^2$ weight of the operator $\p_7$ in
Table~\ref{table:operators},
$$ \delta \kappa_V^2 \simeq \frac{1}{2\sqrt{2} G_F}\frac{m_h^2}{m_\rho^4}\,.$$

\subsubsection*{Fermionic sector}

Consider first the case in which the fermion mass is generated by a
single Yukawa operator $\mathcal{O}_{{\rm Yuk},f}^{(n,m)}$, see
Eqs.~(\ref{eq:generic_yukawa_Lag}) and (\ref{eq:generic_yukawa}).
From the Lagrangian in Eqs.~(\ref{L_expanded})-(\ref{eq:L_2}), and
more specifically from the Yukawa operator in the third line of
Table~\ref{table:operators}, an expression for the fermion mass follows
after applying Eqs.~(\ref{vevs-polar}) and Eq.~(\ref{eq:Zh}),
\begin{equation}\label{eq:m_f}
 \mathcal{L}_{\rm{f}}^{\rm{Yuk}}\supset -m_{{\rm f}}\, \bar{{\rm f}}_L {\rm f}_R+{\rm h.c.}\,,     \qquad m_{{\rm f}}\simeq \frac{y_{{\rm f}}^0}{\sqrt{2}}\,f\,\sqrt{\xi}\,(1-\xi)^{n/2}\xi^m\left(1+n\, \frac{1}{\xi(1-\xi)}\,\frac{m_h^2}{m_\rho^2}\right)\,.
\end{equation}
{\bf Renormalization.} The renormalization scheme is now enlarged to the fermion sector 
choosing as observable precisely the
fermion masses. The prediction that follows for  the
Higgs coupling to a given fermion ${\rm f}$,
\begin{equation}
  \mathcal{L}_{h{{\rm f}}{{\rm f}}} \equiv - g_{h{{\rm f}}{{\rm f}}} \, h \, {\bar{{\rm f}}_L}  {\rm f}_R+{\rm h.c.}\,,
\end{equation}
takes then the form 
\begin{align}\label{eq:g_hff}
  g_{h{{\rm f}}{{\rm f}}}\simeq \frac{y_{{\rm f}}^0}{\sqrt{2}} &(1-\xi)^{\frac{n-1}{2}}\xi^m\Big\{
  (1+2m)(1-\xi)-n\xi +\frac{\beta}{\xi(1-\xi)}\frac{m_h^2}{m_\rho^2}\times\\
  &\times\big[ (1+2m+n)\xi(1-\xi)(2-\xi) + n\left(1+2m(1-\xi)-n\xi\right) \big]
 \Big\}\,.\nonumber
\end{align}
Encoding the deviations with respect to the SM expectations through
the conventional $\kappa_{{\rm f}}$ parameter,
\begin{equation}
\kappa_{{\rm f}}\equiv g_{h{{\rm f}}{{\rm f}}}/g_{h{{\rm f}}{{\rm f}}}^{\rm SM}\,, 
\end{equation}
where $g_{h{{\rm f}}{{\rm f}}}^{\rm SM}=m_{{\rm f}}/v$, the exact and
somewhat lengthy expression for $\kappa_{{\rm f}}$ up to order
$1/\lambda$ follows.  The latter can be simply recast assuming again
$\mathcal{O}(\xi)\sim \mathcal{O}(1/\lambda)$, leading to
\begin{equation}\label{eq:kappa_f}
  \kappa_{{\rm f}}\simeq \frac{(1+2m)(1-\xi)-n\xi}{\sqrt{1-\xi}}+(2+4m+3n)\,\frac{m_h^2}{m_\rho^2},
\end{equation}
where once again Eq.~(\ref{linear-correction}) has been used. 
It is straightforward to check that the first term on the right-hand
side of this equation reproduces well-known $\kappa_{{\rm f}}$ results
for different models in the literature, which assume a non-linear
realization. The second term gives instead the leading linear
corrections. For instance, this equation leads to the
following results for the  MCHM$_{5-1-1}$ (corresponding to $n=m=0$
in our parametrization) and MCHM$_5$ (corresponding to $n=1$, $m=0$):
 \begin{equation}\label{eq:kappa_f_4_5}
  \kappa_{{\rm f}}^{{\rm MCHM}_{5-1-1}}\simeq\sqrt{1-\xi}+2\frac{m_h^2}{m_\rho^2},\quad
\kappa_{{\rm f}}^{{\rm MCHM}_5}\simeq\frac{1-2\xi}{\sqrt{1-\xi}}+5\frac{m_h^2}{m_\rho^2}\,.
\end{equation}
obtaining again at order $1/\lambda$ a correction doubly suppressed  as proportional to both $\beta$ and $\xi$,  see Eq.~(\ref{linear-correction}).

Consider next the case in which a given fermion mass corresponds to
the combination of several $SO(5)$ invariant Yukawa operators, instead
of just one as developed above,
\begin{equation}\label{eq:generic_yukawa_Lag_nm}
  \mathcal{L}_{{\rm f}}^\text{Yuk} =-\,c_{(n,m)}\, \mathcal{O}_{{\rm Yuk},{{\rm f}}}^{(n,m)}
  + \dots + {\rm h.c.}, 
  \end{equation}
where $c_{(n,m)}$ are related to the generators of $SO(5)$ and
the fermion embedding in a given model. The procedure is still quite
straightforward. The fermion mass will be a sum of contributions
similar to that in Eq.~(\ref{eq:m_f}) weighted by the coefficients
$c_{(n,m)}$, and a similar combination protocol will apply to the
obtention of the fermion-Higgs coupling $g_{h{{\rm f}}{{\rm f}}}$ and
$\kappa_{{\rm f}}$.  As an example, consider the MCHM$_{14-14-10}$
scenario~\cite{Carena:2014ria}, in which the third family quark doublet and the right-handed
top are embedded each in a $\mathbf{14}$-plet of $SO(5)$, denoted
$Q_L$ and $U_R$ respectively, while the right-handed bottom is
included in a $\mathbf{10}$-plet representation denoted $D_R$. Two
$SO(5)$ invariant operators~\cite{Carena:2014ria} contribute in this
case to the top quark mass,
\begin{equation}
  y_u\phi^\dagger \bar{Q}_LU_R\phi - \tilde{y}_u(\phi^\dagger\bar{Q}_L\phi)(\phi^\dagger U_R\phi)
\to 3y_u\mathcal{O}_{\rm Yuk}^{(1,0)} - \tilde{y}_u\left(2\mathcal{O}_{\rm Yuk}^{(1,1)} -8\mathcal{O}_{\rm Yuk}^{(3,0)}\right)\,,
\end{equation}
leading to
\begin{equation}
  \kappa_{\rm t}^{{\rm{MCHM}}_{14-14-10}}\simeq\frac{y_u(3-6\xi)+2\tilde{y}_u(4-23\xi+20\xi^2)}{\sqrt{1-\xi}\left(3y_u+2\tilde{y}_u(4-5\xi)\right)}
+\frac{15y_u^2+32\tilde{y}_u(8\tilde{y}_u-3y_u)}{(8\tilde{y}_u-3y_u)^2}\frac{3m_h^2}{m_\rho^2}\,.
\end{equation}
 In contrast, in this
same scenario only one effective Yukawa operator contributes to the
bottom quark mass,
\begin{equation}
  y_d\phi^\dagger \bar{Q}_LD_R\phi 
\to y_d\mathcal{O}_{\rm Yuk}^{(1,0)} \,,
\end{equation}
and consequently 
\begin{equation}
\kappa_{\rm b}^{{\rm{MCHM}}_{14-14-10}}\simeq  \frac{1-2\xi}{\sqrt{1-\xi}}+\frac{5\beta \xi}{4\lambda} \simeq \frac{1-2\xi}{\sqrt{1-\xi}}+\frac{5m_h^2}{m_\rho^2} \,.
\end{equation}
All $\mathcal{O}(1/\lambda)$  corrections considered above show again the double suppression in $\xi$ and $\beta$, which after Eq.~(\ref{linear-correction}) is tantamount to a $m_h^2/m_\rho^2$ suppression factor, as expected. 

\section{Explicit fermion sector}
\label{explicit-heavy-fermions}
In the previous section, the infinite mass limit for the heavy fermion
sector was assumed from the start, while the corrections due to the heavy scalar
singlets were explored.  In this section we start instead of a
complete (bosons plus fermions) renormalizable model, so as to
estimate the impact of a fermionic ultraviolet completion beyond that
related to the Yukawa couplings discussed earlier. The low-energy
effective Lagrangian made out of SM fields will be then explicitly determined up
to the leading corrections stemming from the heavy scalar and fermion
sectors: respectively up to $\mathcal{O}(1/\lambda)\sim
\mathcal{O}(m_h^2/m_{\rho}^2)$ and $\mathcal{O}(f/\mathcal{M}_i)$,
where $\mathcal{M}_i$ denotes generically the heavy fermion masses.

The details
of the fermion mass Lagrangian are quite model-dependent and derived
from the specific $SO(5)$ embedding of the light and heavy fermions.
Many choices of fermion representations are possible.  In addition to the use of heavy vectorial representations, a common trend
is to avoid by construction direct Yukawa couplings of the Higgs field
to the SM fermions, 
leading to a generalized see-saw pattern for light fermions with
masses inversely proportional to those for the heavy fermions.

The fermionic Lagrangian $\mathcal{L}_{\rm f}$ in Eq.~(\ref{eq:Complete_Lagrangian}) needs to be redefined,
\begin{equation}\label{eq:Fermion_Lagrangian-explicit-fermions}
 \mathcal{L}_{{\rm f}}= \mathcal{L}_{{{\rm f}}}^{\rm kin} + \mathcal{L}_{{\rm f}}^{\rm Yuk}\,,
\end{equation} 
where  $\mathcal{L}_{{{\rm f}}}^{\rm kin}$ contains now kinetic terms for all fermions, light and heavy, and the fermion mass Lagrangian denoted by $\mathcal{L}_{{\rm f}}^{\rm Yuk}$ needs to be specified for a particular ultraviolet fermion completion.
\label{eq:Fermion_Lagrangian}
The model developed in Ref.~\cite{Feruglio:2016zvt} will be analyzed
as illustration, recalling first its main ingredients.  In order to
obtain the correct hypercharge assignments, the symmetry of the
Lagrangian needs to be enlarged as customary to $SO(5) \times U(1)_X$
which is broken down to $SU(2)_L \times SU(2)_R \times U(1)_X$, where
the hypercharge corresponds to $Y = \Sigma^{(3)}_R + X$. The fermion
fields that will generate the top mass are
\begin{align}
  &\psi^{(2/3)} \sim (X,Q,T^{(5)}) \sim (\mathbf{2}_{+7/6},\mathbf{2}_{+1/6},\mathbf{1}_{+2/3}), 
  \quad\quad\, \chi^{(2/3)} \sim T^{(1)}\sim (\mathbf{1}_{+2/3}), \nn \\
  &\psi^{(-1/3)} \sim(Q',X',B^{(5)})\sim (\mathbf{2}_{+1/6},\mathbf{2}_{-5/6},\mathbf{1}_{-1/3}),
\quad \chi^{(-1/3)} \sim B^{(1)}\sim(\mathbf{1}_{-1/3}),\nn 
\end{align}
where $\psi^{(x)}$ and $\chi^{(x)}$ belong respectively to the
$\mathbf{5}$ and $\mathbf{1}$ representations of $SO(5)$ with $U(1)_X$
charge $x$; their decomposition in terms of $SU(2)_L\times U(1)_Y$
charges is also shown. This choice of heavy fermion representations corresponds to the MCHM$_{5-1-1}$ scenario, that is, to the entry $5-1$ in the first row of Table~\ref{table:yukawas}, and thus to the effective Yukawa operator 
$\mathcal{O}_{{\rm Yuk},\rm{f}}^{(n,m)}$ in Eq.~(\ref{eq:generic_yukawa})  with $\{n,m\}=\{0,0\}$.

 The fermionic Lagrangian for that field
content reads
\begin{align}
\mathcal{L}_{{\rm f}}
   &= \bar{q}_{L} i\Ds  q_L + \bar{t}_R i\Ds  t_R + \bar{b}_R i\Ds  b_R\nn\\ 
  & + \bar{\psi}^{(2/3)}\left(i\Ds-M_5\right)\psi^{(2/3)}+\bar{\psi}^{(-1/3)}\left(i\Ds-M'_5\right)\psi^{(-1/3)}\nn \\
   &+ \bar{\chi}^{(2/3)}\left(i\Ds-M_1\right)\chi^{(2/3)}+\bar{\chi}^{(-1/3)}\left(i\Ds-M'_1\right)\chi^{(-1/3)}\nn \\
   &-\Big[ y_1\,\bar{\psi}_{L}^{(2/3)} \phi \,\chi_{R}^{(2/3)}+y_2\,\bar{\psi}_{R}^{(2/3)} \phi \,\chi_{L}^{(2/3)} 
    +y'_1\,\bar{\psi}_{L}^{(-1/3)} \phi \,\chi_{R}^{(-1/3)}+y'_2\,\bar{\psi}_{R}^{(-1/3)} \phi \,\chi_{L}^{(-1/3)}\nn \\
   &+ \Lam_1\paren{\bar{q}_L{\De_{2\times5}^{(2/3)}}}\psi_R^{(2/3)}		          						      
    + \Lam_2 \,\bar{\psi}_L^{(2/3)} \paren{\De_{5\times1}^{(2/3)} t_R} + \Lam_3 \,\bar{\chi}_L^{(2/3)} t_R \nn \\
   &+ \Lam_1'\paren{\bar{q}_L{\De_{2\times5}^{(-1/3)}}}\psi_R^{(-1/3)}   
     + \Lam'_2 \,\bar{\psi}_L^{(-1/3)} \paren{\De_{5\times1}^{(-1/3)} b_R} +\Lam'_3 \,\bar{\chi}_L^{(-1/3)} b_R+h.c.\Big] \,,
\label{SO5Lag}
\end{align}
where $\Delta^{(x)}_{n\times m}$ are spurion fields that break
explicitly $SO(5)$, while all other terms are $SO(5)$ invariant. In
terms of $SU(2)_L$ fields, the Yukawa and spurion terms in
Eq.~(\ref{SO5Lag}) provide now an explicit realization of the
Lagrangian $\mathcal{L}_{{\rm f}}^{{\rm Yuk}}$ in
Eq.~(\ref{eq:Fermion_Lagrangian-explicit-fermions}):
\begin{align}
\mathcal{L}_{{\rm f}}^{\rm Yuk}
    =&-\Big[y_1 \paren{\bar{X}_L H{}  T_R^{(1)} +\bar{Q}_L \widetilde{H}{}  T_R^{(1)}
        +\bar{T}_L^{(5)}\sigma T_R^{(1)}}\label{eq:yukawa-heavy-fermionsSO5LagD}\\
  &+y_2\,\paren{\bar{T}_L^{(1)} H{}^\dagger X_R 
        +\bar{T}_L^{(1)}\widetilde{H}{}^\dagger Q_R+\bar{T}_L^{(1)}\sigma T_R^{(5)}} \nn \\
    &+y'_1 \paren{\bar{X'}_L \widetilde{H} B_R^{(1)} +\bar{Q'}_L H B_R^{(1)}
        +\bar{B}_L^{(5)}\sigma B_R^{(1)}} \nn\\
&+ y'_2\,\paren{\bar{B}_L^{(1)} \widetilde{H}^\dagger X'_R 
        +\bar{B}_L^{(1)} H^\dagger Q'_R+\bar{B}_L^{(1)}\sigma B_R^{(5)}} \nn \\
    &+\Lam_1\bar{q}_L Q_R+ \Lam_1'\bar{q}_L Q_R'+\Lam_2\bar{T}_L^{(5)} t_R+\Lam_3\bar{T}_L^{(1)} t_R 
      +\Lam'_2\bar{B}_L^{(5)} b_R+\Lam'_3\bar{B}_L^{(1)} b_R + h.c.\Big].\nn
\end{align}
In Ref.~\cite{Feruglio:2016zvt} we had first integrated out the heavy fermions of this Lagrangian, determining then the effective Lagrangian made out of SM fields plus the singlet scalar present in the minimal $SO(5)$ sigma model. Here we reverse the order of integration of the heavy fields, taking first the limit of heavy $\rho$ and then that of heavy BSM fermions. We have explicitly
checked that the final  low-energy effective Lagrangian made out only of SM fields is independent of the order in which
those limits are taken. 

Using polar coordinates and integrating out the radial mode $\rho$
does not bring any novel complication with respect to the procedure
carried out in the previous section, except for lengthier expressions.
Nevertheless, $\mathcal{L}_F^{\rm Yuk}$ can be compactly written prior to any integration procedure as
\begin{align}
\mathcal{L}_F^{\rm Yuk}
    =-\Big[&\rho \left(s_\varphi \mathcal{O}^F_s + c_\varphi \mathcal{O}^F_c\right)+\Lam_1\bar{q}_L Q_R+ \Lam_1'\bar{q}_L Q_R'\nn\\
&+\Lam_2\bar{T}_L^{(5)} t_R+\Lam_3\bar{T}_L^{(1)} t_R +\Lam'_2\bar{B}_L^{(5)} b_R+\Lam'_3\bar{B}_L^{(1)} b_R + h.c.\Big]\,,
\end{align}
where $\mathcal{O}^F_s$ and $\mathcal{O}^F_c$ are heavy fermion bilinears corresponding to the first four lines in Eq.~(\ref{eq:yukawa-heavy-fermionsSO5LagD}):  
\begin{align}
&\mathcal{O}^F_s\equiv  -\frac{1}{\sqrt{2}}\Big[y_1\! \paren{\bar{X}_L U e_-  T_R^{(1)} +\bar{Q}_L U e_+  T_R^{(1)}}
    +y_2\paren{\bar{T}_L^{(1)} U e_- X_R+\bar{T}_L^{(1)} U e_+ Q_R}\nn\\
    &\quad+y'_1\paren{\bar{X}'_L U e_+ B_R^{(1)} +\bar{Q}'_L U e_- B_R^{(1)}}
    + y'_2\paren{\bar{B}_L^{(1)} U e_+ X'_R+\bar{B}_L^{(1)} U e_- Q'_R}\Big],\\
&\mathcal{O}^F_c\equiv  -\frac{1}{\sqrt{2}}\Big[y_1\bar{T}_L^{(5)} T_R^{(1)}+y_2\bar{T}_L^{(1)} T_R^{(5)}
    +y'_1 \bar{B}_L^{(5)} B_R^{(1)}+ y'_2\bar{B}_L^{(1)} B_R^{(5)}\Big]\,,
\end{align}
where $e_+ = (1,0)$ and $e_-=(0,1)$.
\begin{table}
  \centering
\renewcommand{\arraystretch}{2.2}
\begingroup\small
\begin{tabular}{|c| c| c|c|}
\hline
& Operator &  $\F_i(\varphi)$& $1/\lambda^n$  \\
\hline
$\p_{\rm Yuk}$ & $v(\bar q_{iL} U P_\pm \qri)$&  
	$-\dfrac{y_t^0}{\sqrt{2\xi}}s_\varphi\left[1-\dfrac{1}{8\lambda}(\alpha c_\varphi -2\beta s_\varphi^2)- 2\dfrac{f}{\mathcal{M}_i}a_{\sigma 1}^ic_\varphi\right]$
   &  0 \\
\hline
$\p_{qh}$ & $\paren{\dmu h}^2(\bar q_{iL} U P_\pm \qri)$&  
	$-\dfrac{y_i^0}{8\sqrt{2}\lambda f^3}\sh \paren{1-2\dfrac{f}{\mathcal{M}_i}a^{i}_{\sigma 1}\ch}$   &  1 \\  
\hline
$\p_{qV}$ & $\VV(\bar q_{iL}UP_\pm \qri)$&  
	$\dfrac{y_i^0}{16\sqrt{2}\lambda f}\sh \paren{1-2\dfrac{f}{\mathcal{M}_i}a^{i}_{\sigma 1}\ch}$   &  1 \\
\hline
$\p_{4q}$ & $\paren{\bar q_{iL}UP_\pm \qri}\paren{\bar q_{jL}UP_\pm \qrj}$&  
	$(2-\delta_{ij})\dfrac{y_i^0 y_j^0}{32 \lambda f^2}\sh^2 \left[1-2
             \left(a^{i}_{\sigma 1}\dfrac{f}{\mathcal{M}_i}+a^{j}_{\sigma 1}\dfrac{f}{\mathcal{M}_j}\right)\ch\right]$&  1 \\
\hline
$\p_{4q'}$ & $\paren{\bar q_{iL}UP_\pm \qri}\paren{\barqrj P_\pm U^\dagger q_{jL}}$&  
	$(2-\delta_{ij})\dfrac{y_i^0 y_j^{0}}{32 \lambda f^2}\sh^2 \left[1-2
             \left(a^{i}_{\sigma 1}\dfrac{f}{\mathcal{M}_i}+a^{j}_{\sigma 1}\dfrac{f}{\mathcal{M}_j}\right)\ch\right]$& 1  \\
\hline
\end{tabular}
\endgroup
\caption{Effective operators, up to order $f/\mathcal{M}_i$ and
  $1/\lambda$, after integrating out the radial mode $\rho$ and the
  heavy fermions in a UV realisation of partial compositeness. The
  coefficients $a_{\sigma 1}^f$, with $f=t,b$ can be found in
  Ref.~\cite{Feruglio:2016zvt}, see footnote 6. The Hermitian
  conjugate should be included for all operators here. The Higgs field $h$ is defined as the
  excitation of the field $\varphi$, see Eq.~(\ref{vevs-polar}).}
\label{table:operators-fermion}  
\end{table}

Consider next the limit of very large scalar mass $m_\rho$ (that is
$\lambda \to \infty$) and very heavy fermions. Implementing first the $1/\lambda$
corrections, the effective Lagrangian at this order takes exactly the
form in Eq.~(\ref{eq:L_1}), although $\rho_1$ shows now an explicit
dependence on the heavy fermion spectrum,
\begin{equation}
\rho_1 = \frac{f}{4}
  \left[\frac{1}{2f^4}\partial_\mu \varphi\,\partial^\mu\varphi
    -\frac{1}{4f^2}\vev{V_\mu V^\mu}s_\varphi^2 - \frac{1}{2}\alpha c_\varphi + \beta s_\varphi^2 -\left\{\frac{1}{2f^3}\mathcal{O}^F_c c_\varphi 
    +\frac{1}{2f^3}\mathcal{O}^F_s s_\varphi \,+\,h.c.\right\}\,\right]\,,\nn
\end{equation}
instead of the effective dependence in Eq.~(\ref{eq:expressions-rho}). New operators  beyond those previously considered appear, such as 
\begin{align}
  &-\frac{1}{8\lambda f^3}\partial_\mu \varphi\,\partial^\mu \varphi 
         \,\left(\mathcal{O}^F_c c_\varphi+\mathcal{O}^F_s s_\varphi\right)\,,\\
  & \frac{1}{16\lambda f}\vev{V_\mu V^\mu} s_\varphi^2 
         \left(\mathcal{O}^F_c c_\varphi+\mathcal{O}^F_s s_\varphi\right)\,,\\
  & \frac{1}{16\lambda f^2}\left(\mathcal{O}^F_c c_\varphi+\mathcal{O}^F_s s_\varphi\right)^2\,.
\end{align}
  They are higher-order operators made out of both SM and heavy BSM fermions  and related to the explicit fermionic
  ultraviolet completion. Furthermore, it is again easy to verify that the counting rule matches the NDA
rule~\cite{Manohar:1983md, Gavela:2016bzc} by identifying $\lambda
f\sim\Lambda$.
 
Consider next the integration of
the heavy fermion sector in the results just obtained.
This is an elaborated task, and the procedure and an explicit computation is described in
Ref.~\cite{Feruglio:2016zvt}.  To estimate the
corrections, we adopt here a universal heavy fermion mass scale
$\mathcal{M}_i$ associated with the mass generation mechanism of a
given SM fermion, so that $M_1\sim M_5\sim
\Lambda_1\sim\dots\sim\mathcal{M}_t$. Assuming this scale to be larger
than $f$, $f/\mathcal{M}_i$ is a good expansion parameter.  The final
set of five effective operators resulting up to  first order in the
$1/\lambda$ and $f/\mathcal{M}_i$ expansions is shown in
Table~\ref{table:operators-fermion}, where the $a_{\sigma 1}^i$
operator coefficients weighting the $f/\mathcal{M}_i$ corrections are
expected to be $\mathcal{O}(1)$ and their exact expressions can be found 
in Ref.~\cite{Feruglio:2016zvt}.~\footnote{The $a_{\sigma 1}^i$
  coefficients are a redefinition of the $c_{\sigma}^i$ operator
  coefficients in Ref.~\cite{Feruglio:2016zvt} so as to extract
  explicitly the $f/\mathcal{M}_i$ dependence: $c_{\sigma 1}^i \to
  y^0_t\,a_{\sigma 1}^t/{\mathcal{M}_t}$; the exact expressions for
  $c_{\sigma}^i$ for the fermion model discussed here can be found in
  Table 3 of that reference.} Noteworthy consequences include:
\begin{itemize}
\item At tree level, the heavy fermions have no impact on the
  gauge-Higgs coupling and $\kappa_V$ is still given by
  Eq.~(\ref{eq:kappa_V}). The coupling to top quarks, on the contrary,
  will receive fermionic contributions from the first operator in
  Table~\ref{table:operators-fermion},
  \begin{equation}
    \kappa_{\rm t}=\sqrt{1-\xi}+2\,\frac{m_h^2}{m_\rho^2}+a_{\sigma
      1}^t\frac{f}{\mathcal{M}_t}\xi+\dots
  \end{equation}
  Again, a double suppression acts on the leading heavy fermion
  corrections $\sim \xi f/\mathcal{M}_t$, alike to the case for the bosonic ones in $\sim\beta\xi/(2\lambda)$. It is important to note,
  though, that the tree-level fermionic contributions found may be larger
  than those induced by the scalar sector if $f/\mathcal{M}_t >
  \beta/\lambda$; this may occur specially for the top quark since the
  top partners, with characteristic mass scale $\mathcal{M}_t$, should
  be light enough in order not to generate a hierarchy problem.
\item On top of the above, higher order effective operators involving
  SM fields are singled out at low scales: the dominant ones are the
  last three presented in Table~\ref{table:operators-fermion}. For
  these operators, the inclusion of an explicit heavy fermion sector
  does not change much the conclusions obtained previously by using an
  effective Yukawa coupling as defined in
  Eq.~(\ref{eq:generic_yukawa}).
\item In the limit $f/\mathcal{M}_i\to 0$, the operators in
  Table~\ref{table:operators-fermion} coincide as expected with the
  fermion-Higgs and four fermion operators given previously in
  Table~\ref{table:operators} using the effective Yukawa operator
  $\mathcal{O}^{(0,0)}_{\rm Yuk}$.
\end{itemize}

\section{Conclusions}
The linear sigma model for QCD allows to monitor the transition from a
completely renormalizable model in a weakly interacting regime to a
non-linear regime in the high mass ($\lambda\to \infty$) limit. In
this work we have carried out an analogous exploration assuming that the
Higgs particle may correspond to a pseudo-Goldstone boson of a
spontaneously $SO(5)$ global symmetry (or containing $SO(5)$) at high
energies, completing the procedure first proposed and started in Ref.~\cite{Feruglio:2016zvt}.

The minimal sigma model for $SO(5)$ has been used as starting
point. The results are independent of the relative order in which the
high mass limit for the heavy boson $\rho$ and for the heavy fermions
are taken.  In a first stage, the bosonic sector was left fully
dynamical while we defined an effective Yukawa operator characterized
by two parameters, which depend only on how the light SM fermion
fields are embedded in representations of the $SO(5)$ symmetry for any
given model. Armed with this tool, the benchmark effective Lagrangian
has been derived for the large $m_\rho$ limit, see Table~\ref{table:operators}.  Up to first order in
the linear corrections, it is shown to be composed of ten operators,
five of them bosonic and the rest fermionic including that responsible
for the usual fermion Yukawa coupling; for the fermionic operators the
coefficients are given as an explicit function of the two  parameters which define 
the effective Yukawa operator.  Their simple form allows a direct prediction and comparison of the many models in the literature which differ by their fermionic embedding. It is straightforward to obtain from this result the expressions of the Higgs couplings to fermions, $\kappa_{\rm{f}}$, in general.

Among the $\mathcal{O}(1/\lambda)$
corrections, three bosonic operators have coefficients which are independent
of the global symmetry breaking mechanism and should thus be of
special relevance; among them a four-gauge boson vertex is already
being directed probed  by present data, while the other two involve vertices with
at least two Higgs fields.  We have also proved that the leading
phenomenological couplings of the Higgs particle to gauge bosons and
SM fermions, $\kappa_V$ and $\kappa_{{\rm f}}$, are quite universal,
the reason being that the linear corrections must be doubly suppressed
as proportional to both $1/\lambda$ and to the explicit symmetry
breaking parameters, in a combination corresponding to a
$m_h^2/m_\rho^2$ suppression.

The tower of higher order operators obtained is shown to correspond
 to only a small subset of the most general non-linear
Lagrangian~\cite{Alonso:2014wta} for a generic non-linear realization
of electroweak symmetry breaking. It is also consistent with the results for general $SO(5)/SO(4)$ constructions in Ref.~\cite{Alonso:2014wta}, singling out a fraction of operators found in the latter, with its expected coefficients. The minimal set identified here could serve to
focalize model-independent searches of a dynamical nature for the
Higgs particle.

In a second stage, we have explored a complete renormalizable model
with an explicit heavy fermion ultraviolet completion and repeated the
integration procedure; the leading corrections stemming both from the
heavy boson and from the heavy fermion sector were then
identified and their coefficients determined.  New higher-order operators related to the specific
fermion ultraviolet completion,  made out of SM and heavy fermions and containing vertices with at least
four fields, were identified as an intermediate step.
Finally,  the set of operators made out of SM fields and involving fermions was determined to consist of only five operators at the
leading order in both expansions: one coupling contains the usual SM
Yukawa coupling, two are fermion-boson operators and the remaining two 
correspond to four fermion couplings. The results match those
obtained in the first part using the effective Yukawa operator;
interestingly, they show that deviations to the SM value for
$\kappa_{{\rm f}}$ due to tree-level exchange of heavy fermions may
dominate over those stemming from the scalar (e.g. $1/\lambda$) linear
corrections. 

The starting point of the analysis in this work is a minimal
renormalizable sigma model for $SO(5)$. Other renormalizable -more
complicated- realizations are conceivable, in the same way that the
linear sigma model for QCD could be extended. Although additional
effective operators could be sourced in such constructions~\cite{Alonso:2014wta}, the
operators identified here are expected to be the tell-tale of a
Goldstone boson origin for the Higgs field and as such common to all
realizations.
 
\section{Acknowledgements}
We specially acknowledge initial discussions with Ferruccio Feruglio and
Stefano Rigolin. We are also indebted to Ilaria Brivio and Luca Merlo
 for useful discussions.  The authors (each identified
by the first letter of her/his last name) acknowledge partial
financial support by the European Union through the FP7 ITN INVISIBLES
(PITN-GA-2011-289442) (GKMS), by the Horizon2020 RISE InvisiblesPlus
690575 (GKMS), by CiCYT through the project FPA2012-31880 (GS), and by
the SpanishMINECO through the Centro de excelencia Severo Ochoa
Program under grant SEV-2012-0249 (GMS).  We
are grateful to the Physics Department of the University of
California, Berkeley, the Lawrence Berkeley National Laboratory (SG)
and the Fermi National Accelerator Laboratory (K), and the
Universidade de S\~{a}o Paulo (M) for hospitality and/or partial
support during the completion of this work. The work of S.S. was 
supported through the grant BES-2013-066480 of the Spanish MICINN. The
work of K.K. is supported by the University of Padova. In the final stages of this paper, the work of P.M. was supported  by Fermilab, which is
operated by the Fermi Research Alliance, LLC under contract
No. DE-AC02-07CH11359 with the United States Department of Energy.

\end{document}